\documentstyle[12pt]{article}
\title{On signature transition in Robertson-Walker cosmologies}
\author{K. Ghafoori-Tabrizi, S. S. Gousheh and H. R. Sepangi\thanks{e-mail: 
hr-sepangi@cc.sbu.ac.ir}\\ {\small Department of Physics, 
Shahid Beheshti University, Evin, Tehran 19839, Iran}}
\begin{document}
\maketitle
\vspace{15mm}
\begin{abstract}
We  analyse  a classical model of gravitation coupled to a self interacting 
scalar field. We show that, within the context of this model for 
Robertson-Walker cosmologies,
there exist  solutions in the spatially non-flat cases exhibiting 
transitions from a Euclidean to a Lorentzian spacetime. We then  discuss  the 
conditions under which these signature changing solutions to Einstein's  field
 equations exist. In particular, we find that an upper bound for the 
cosmological constant exists and that close to  the signature changing 
hypersurface, both the scale factor and the scalar field have to be constant. 
Moreover we find that the signature changing solutions do not exist when the 
scalar field is massless.
\end{abstract}\vspace{10mm}
\section{Introduction}
The principle of causality is one of the most fundamental
pillars on which classical as well as quantum physics is based. 
It exhibits itself in terms of the causal structure of a
four-dimensional Lorentzian space-time manifold. Theories like quantum
fields rely  on this structure  for a sensible interpretation of their
results in a fixed background geometry. However, there is no  {\it apriori}
reason which would suggest that this causal structure should  
remain unaffected by the dynamical
equations of general relativity. If we accept that in quantum cosmology
\cite{Dewitt}, amplitudes for gravity can be expressed as sum of histories
of 3-geometries \cite{Hartle} with different global 4-topological structures,
then this may give rise to 
structures that cannot have pure Lorentzian or Euclidean geometries 
\cite{kundt} and this implies a possible signature transition \cite{HM}.
For example, an interesting feature of present day  quantum cosmology
\cite{Hartle}
is characterised by its use  of Riemannian signature spaces to arrive at
an observable Lorentzian-signature spacetime. Therefore, in order to 
gain further isight, it might be useful 
to study the dynamics of this phenomenon in simple models.   

Traditionally, one of the features of  classical gravity is that the signature
of the metric is usually considered as fixed. This is not a property demanded
by  the field equations, but rather a condition one imposes on the metric
before looking for solutions to Einstein's equations. If one relaxes this
condition, one may find solutions to the field equations, which when
interpreted suitably, exhibit a signature transition. 

In general, for studying signature dynamics in the classical realm of 
gravity \cite{Sakharov}, one assumes that there exists a
hypersurface which separates the space into two disjoint
regions, one being Euclidean and the other Lorentzian. On this 
hypersurface, henceforth called the junction, the metric is either continuous 
and necessarily degenerate, or discontinuous with nondegenerate one-sided 
limits \cite{kkd}. Therefore one can categorize this problem
according to the continuity of the metric. Having chosen the form
of the metric, there are two approaches for solving the field equations.
The first (continuous) approach is to assume the validity of 
Einstein's field equations throughout and identify the form of
the field equations at the junction with the junction conditions 
\cite{Dereli,Hayward,sepangi}. In the second (discontinuous) approach one 
finds solutions by solving Einstein's equations in disjoint regions next 
to the junction and then uses specific junction conditions ({\it i.e.} 
Darmois junction conditions) to match the solutions \cite{Heldray}. This 
furnishes an alternative way of categorizing the problem.

These different approaches  have  resulted in mainly two different sets  of
junction conditions which have led to a controversy between various authors
\cite{heyhel}. Within the framework of Robertson-Walker cosmology using 
either approach,  the scale
factor, the scalar field and their first time derivatives are required to
be continuous across the junction. However 
within the continuous approach, Hayward using a continuous metric has 
obtained  extra conditions restricting the first time derivatives
to be zero at the junction. The controversy mentioned above stems from the
claims by some authors that these extra junction conditions are unnecessary,
see e.g.  \cite{Ellis,NC} and therefore  the matter conservation need not
hold under  signature change \cite{NC}.   

Recently, Kossowski and Kriele \cite{kk}, within the continuous approach, 
have proved a local existence and
uniqueness theorem for solutions of Einstein's equations with dust energy
momentum tensor in a class of m-dimensional signature changing  spacetimes
where initial conditions are given on the hypersurface of signature change.
They have also proved a similar theorem in the case where the energy-momentum
tensor represents a  scalar field which is not self-interacting.

In this paper, we study a classical model where signature transition is 
possible, using the continuous approach with a continuous metric.  This
model is the same as that proposed by Dereli and Tucker \cite{Dereli} in 
which a real massive  scalar field is taken as the matter source interacting 
with gravity and itself in a Robertson-Walker geometry whose signature 
evolution is controlled by a preferred coordinate. One then seeks solutions 
which are smooth and continuous across the junction on which the metric is
degenerate. 
For the spatially flat universes, this results in exactly solvable Einstein's
field equations 
\cite{Dereli}. Here we show that signature changing solutions also exist for
arbitrary curvature in Robertson-Walker cosmologies. We do this by finding
analytic solutions to the field equations close to the junction, and confirm 
and extend them away from the junction by a numerical method.  We also
show systematically the general set of restrictions that need to be imposed
on the junction conditions and the parameters of this model including the 
cosmological constant. The general set of allowed conditions and parameters
thus obtained include the ones  used in reference \cite{Dereli} as a special 
case. In particular we find an upper bound on the cosmological constant, and
the need for the scalar field to be massive in order to have continuous
solutions.
The present work is also complementary to \cite{kk} in that it solves 
the problem for a self-interacting scalar field. 
Having chosen a continuous approach, we obtain a set of junction conditions
which, not surprisingly, are more restrictive than those obtained 
in the discontinuous approach. However, 
the restrictiveness of the junction conditions obtained here lies somewhere
in between those obtained in \cite{Hayward} and \cite{kk}. We shall return 
to this point in the conclusions.

\section{Field equations}
We begin by briefly reviewing the model, and for comparison purposes, use the
same notation  as that of  \cite{Dereli}. 
Consider gravity coupling to a scalar field through
\begin{eqnarray}
G=\kappa T[\phi],  \label{eq1}
\end{eqnarray}
where the scalar field $\phi$ is a solution of
\begin{eqnarray}
\Delta\phi-\frac{\partial U}{\partial\phi}=0. \label{eq2}
\end{eqnarray}
Here, $G=\mbox{Ric}-\frac{1}{2}g{\cal R}$ is the Einstein tensor constructed
 from torsion-free connections compatible with the metric, and $U$ is a scalar
 potential for the real scalar field $\phi$ interacting with itself and
 gravity through the stress-energy tensor $T[\phi]$, given by
\begin{eqnarray}
T[\phi]=d\phi\otimes d\phi-\frac{1}{2}g(d\phi,d\phi)g-U(\phi)g. \label{eq3}
\end{eqnarray}  
The above coupled equations are to be solved in a domain that would lead to
 Robertson-Walker cosmologies with Lorentzian signature. However, if the
 metric is suitably parametrized, one can expect to see continuous transition
 to a Euclidean domain. As in \cite{Dereli}, we adopt a chart with coordinate
 functions $\{\beta,x^1,x^2,x^3\}$ where the hypersurface of signature change
 would be characterised by $\beta=0$. The metric can be parametrized to  take
 the form
\begin{eqnarray}
g=-\beta d\beta\otimes d\beta+\frac{R^2(\beta)}{[1+(k/4)r^2]^2}\sum_i dx^i
\otimes dx^i,  \label{eq4}
\end{eqnarray}
where $r^2=\sum x_i x^i$, and depending on the values of  $k=\{-1,0,1\}$ one
 would have  an open, flat or closed universe, respectively. Now, it is
 apparent that the sign of $\beta$ determines the geometry, being Lorentzian
 if $\beta>0$ and Euclidean if $\beta<0$. For $\beta>0$, the traditional
 cosmic time can be recovered by the substitution  $t=(2/3)\beta^{3/2}$.
 Adopting  the chart $\{t,x^i\}$ and using equations (\ref{eq1}) through 
(\ref{eq4}) with units in which $\kappa=1$, one finds
\begin{eqnarray}
3\left[\left(\frac{\dot{R}}{R}\right)^2+\frac{k}{R^2}\right]&=
&\frac{\dot{\phi}^2}{2}+U(\phi), \label{eq5}\\
2\left(\frac{\dot{R}}{R}\right)^{\! \mbox{\normalsize  .}}+
3\left(\frac{\dot{R}}{R}\right)^2+\frac{k}{R^2}&=
&-\frac{\dot{\phi}^2}{2}+U(\phi), \label{eq6}\\
\ddot{\phi}+3\frac{\dot{R}}{R}\dot{\phi}+\frac{\partial U}{\partial\phi}&=
&0, \label{eq7}
\end{eqnarray}
where a dot represents differentiation with respect to $t$. These equations
 are not all independent. Equation (\ref{eq6}) can be obtained by combining
 equations (\ref{eq5}) and (\ref{eq7}). Equations (\ref{eq5}) through
 (\ref{eq7}) can also be derived in the context of Lagrangian dynamics.
 One defines the action 4-form
\begin{eqnarray}
\Lambda=\left[\frac{1}{2}{\cal R}-\frac{1}{2}g(d\phi,d\phi)-U(\phi)\right]\ast
 1, \label{eq7.1}
\end{eqnarray}
whose variations with respect to $g$ and $\phi$ give equations (\ref{eq1})
 and (\ref{eq2}), respectively. This action  can be written as 
\begin{equation}
\Lambda={\cal L}dt\wedge dx^1\wedge dx^2\wedge dx^3, \label{eq7.2}
\end{equation}
where 
\begin{equation}
{\cal L}dt=\left\{-3R\dot{R}^2+3kR+R^3[\dot{\phi}^2/2-U(\phi)]\right\}dt+
d(3R^2\dot{R}). \label{eq7.3}
\end{equation}
After dropping the total derivative term and using the following
 transformation,
\begin{eqnarray}
X=R^{3/2}\cosh(\alpha\phi), \label{eq8}\\
Y=R^{3/2}\sinh(\alpha\phi), \label{eq9}
\end{eqnarray}
we obtain a mechanical analogue for this system given by
the Lagrangian
\begin{eqnarray}
2\alpha^2{\cal L}dt=\left\{-\dot{X}^2+\dot{Y}^2+\frac{9k}{4}(X^2-Y^2)^{1/3}
-2\alpha^2(X^2-Y^2)U(\phi(X,Y))\right\}dt, \label{eq10}
\end{eqnarray}
with $-\infty<\phi<\infty$, $0\le R<\infty$ and $\alpha^2=\frac{3}{8}$. By a 
straightforward Legendre transformation we obtain the ``Hamiltonian"
\begin{eqnarray}
2\alpha^2{\cal H}dt=\left\{-\dot{X}^2+\dot{Y}^2-\frac{9k}{4}(X^2-Y^2)^{1/3}+
2\alpha^2(X^2-Y^2)U(\phi(X,Y))\right\}dt, \label{eq7.4}
\end{eqnarray}
which we identify with the total energy density of the system. If we transform 
this expression back to the form of $R$ and $\phi$, we see that equation
 (\ref{eq5}) is nothing more than a ``zero energy condition." Of course any
 solution of the Euler-Lagrange equations would yield a constant total energy.
 However, Einstein's equations demand zero energy solutions only. 

The above Lagrangian is considerably simplified if we take the potential to be
\begin{eqnarray}
2\alpha^2(X^2-Y^2)U(\phi(X,Y))=a_1X^2+a_2Y^2+2bXY, \label{eq11}
\end{eqnarray}
where $a_1$, $a_2$ and $b$ are free parameters. This potential is required to
 have natural characteristics for small $\phi$, so that we may identify the
 coefficient of $\phi^2/2$ in its Taylor expansion as a positive $m^2$ and 
$U(\phi=0)$ as a cosmological constant $\lambda$. In terms of the above
 parameters,  $\lambda=U|_{\phi=0}=a_1/(2\alpha^2)$ and  $m^2=\partial^2 U/
\partial\phi^2|_{\phi=0}=a_1+a_2$. The features of this potential have been
discussed in more detail in \cite{Dereli}.

The dynamical equations for $X$ and $Y$ in terms of the evolution variable
 $\beta$ now become
\begin{eqnarray}
Y^{\prime\prime}=\frac{1}{2}\left(\frac{1}{\beta}\right)Y^\prime-\frac{3}{4}
\beta kY(X^2-Y^2)^{-2/3}-\beta(a_2Y+bX), \label{eq12}\\
X^{\prime\prime}=\frac{1}{2}\left(\frac{1}{\beta}\right)X^\prime-\frac{3}{4}
\beta kX(X^2-Y^2)^{-2/3}+\beta(a_1X+bY), \label{eq13}
\end{eqnarray}
subject to the ``zero energy condition"
\begin{eqnarray}
2\alpha^2{\cal H}=\left(\frac{1}{\beta}\right)(-X^{\prime^{\,2}}+Y^{\prime^
{\,2}})-\frac{9}{4}k(X^2-Y^2)^{1/3}+(a_1X^2+a_2Y^2+2bXY)=0. \label{eq14}
\end{eqnarray}
Here, a prime represents differentiation with respect to $\beta$. 
The coupled equations (\ref{eq12}) and (\ref{eq13}) must now be solved and, 
since their solutions render $\cal H$ a constant in $\beta $,
equation (\ref{eq14}) becomes only a restriction on the initial conditions. 
In the next section we first find analytic solutions to these equations 
which are valid close to the junction, and then find the solutions for the 
full range of $\beta $ by a numerical method.

\section{Solutions}
\subsection{Analytic solutions close to the junction}
In this section we find analytic solutions which are 
valid near the junction, and obtain the restrictions imposed by  equations 
(\ref{eq12})--(\ref{eq14}) on 
the junction conditions. This is done by noting that in order to have well 
behaved  solutions close to $\beta=0$, the first term of equation (\ref{eq14}) 
shows that  we must either have $X^\prime(\beta)\sim \beta^{n_x}$ and 
$Y^\prime(\beta)\sim \beta^{n_y}$, where $n_x$, $n_y\ge 1/2$, or $|X^
\prime(0)|=|Y^\prime(0)|$. However, the first terms on the right hand side 
of equations 
(\ref{eq12}) and  (\ref{eq13})  impose a more severe restriction. These two 
equations admit solutions $X^\prime(\beta)\sim \beta^{1/2}$ and $Y^\prime
(\beta)\sim \beta^{1/2}$ close to $\beta=0$, however, these class of solutions 
do not admit real or $C^2$ solutions across $\beta=0$.
One can show that regular solutions close to $\beta=0$ are of the form 
\begin{eqnarray}
X(\beta) &=& A_x \beta^3+X_0 \hspace{5mm} \mbox{where} \hspace{5mm}  A_x=
\frac{2}{9}
\left[-\frac{3}{4}\frac{kX_0}{(X^2_0-Y^2_0)^{2/3}}+a_1 X_0+bY_0\right], 
\label{eqxc0}\\
Y(\beta) &=& A_y \beta^3+Y_0 \hspace{5mm} \mbox{where} \hspace{5mm}  A_y=
\frac{2}{9}
\left[-\frac{3}{4}\frac{kY_0}{(X^2_0-Y^2_0)^{2/3}}-a_2 Y_0-bX_0\right], 
\label{eqyc0}
\end{eqnarray}
with $X_0\equiv X(0)$ etc. Therefore,  the initial conditions on the first and 
second derivatives must satisfy the relations
\begin{eqnarray}
X^\prime(0)=Y^\prime(0)=0\hspace{3mm} \mbox{and}\hspace{3mm}X^{\prime\prime}
(0)=Y^{\prime\prime}(0)=0.\label{eq15}
\end{eqnarray}

Strictly speaking the conditions on the second derivatives are not initial 
conditions but rather consistency checks, since we have coupled second order 
equations.
The above relations confine  the possible solutions to a restricted class.  
Therefore, the initial values for the functions $X$ and $Y$ must now satisfy, 
c.f. equation (\ref{eq14})
\begin{eqnarray}
-\frac{9}{4}k(X^2_0-Y^2_0)^{1/3}+(a_1X^2_0+a_2Y^2_0+2bX_0Y_0)=0. \label{eq16}
\end{eqnarray}
Although equation (\ref{eq16}) is equivalent to a sixth order algebraic 
equation which cannot be directly solved  analytically, we can solve it by 
going back to the original variables $R$ and $\phi$. The solutions are
either $R(0)=0$ giving $\phi(0)=\pm\infty$, which we exclude because we have 
been seeking continuous solutions across $\beta=0$, or $R(0)\ne 0$ (it is a 
free parameter) with
\begin{eqnarray}
\phi(0)=\frac{1}{2\alpha}\cosh^{-1}\left[\frac{DB\pm b\sqrt{D^2-B^2+b^2}}{B^2-
b^2}\right],
\label{eq166}
\end{eqnarray}
where 
\begin{eqnarray}
D=\frac{9k}{4R(0)^2}-\frac{a_1-a_2}{2}\hspace{5mm}\mbox{and}\hspace{5mm}B=
\frac{a_1+a_2}{2}=\frac{m^2}{2}.\nonumber
\end{eqnarray}
The contour plots of equation (\ref{eq16}) for $k=\pm1$ are  given in figure 1. 
Along the contours, one finds the possible initial values for $X$ and $Y$. Of 
course, the acceptable values of $X$ and $Y$ ( $X>|Y|$) can also be obtained 
analytically from equation (\ref{eq166}).

In order for this model to have a valid quantum extension, at least in the 
scalar sector, we need to require the potential to have a minimum so that 
a real mass can be defined. That is 
\begin{eqnarray}
m_s^2\equiv\left. \frac{\partial^2 U}{\partial\phi^2}\right|_{\phi=\phi_{min}}=
m^2\sqrt{1-
\frac{4b^2}{m^4}}\hspace{5mm}\mbox{where}\hspace{5mm}\phi_{min}=-\frac{1}{2
\alpha}\tanh^{-1}\frac{2b}{m^2}.\nonumber
\end{eqnarray}
Therefore we need to require $ m^2 \ge 2|b| $. This requirement along with 
the requirement of having real values of $X(0)$ and $Y(0)$ satisfying the zero 
energy condition and leading to real values for $R(0)(>0)$ and $\phi(0)$ give 
the following restriction on the parameters 
\begin{eqnarray}
a_1-\frac{9k}{4R(0)^2}\le |b|\le \frac{a_1+a_2}{2}.\label{eq1666}
\end{eqnarray} 
It is important to note here that equations (\ref{eqxc0},\ref{eqyc0})  imply  
\begin{eqnarray}
\dot{X}(0)&=&0\hspace{9mm}\mbox{and}\hspace{3mm}\dot{Y}(0)=0, \label{eqxdot} \\
\ddot{X}(0)&=&\frac{9}{2}A_x\hspace{3mm}\mbox{and}\hspace{3mm}\ddot{Y}(0)=
\frac{9}{2}A_y \label{eqxddot}.
\end{eqnarray}
These in turn lead to
\begin{eqnarray}
\dot{R}(0)&=&0\hspace{10mm}\mbox{and}\hspace{3mm}\dot{\phi}(0)=0, 
\label{eqdot} \\
\ddot{R}(0)&=&\frac{k}{R(0)}\hspace{3mm}\mbox{and}\hspace{3mm}
\ddot{\phi}(0)=-\frac{m^2}{2\alpha}\sinh[2\alpha\phi(0)]-\frac{b}{\alpha}
\cosh[2\alpha\phi(0)]. \label{eqddot}
\end{eqnarray}
The physical significance of these constraints and their relevance to the 
present controversy on the junction conditions will be discussed in the last 
section.

\subsection{Numerical solution}
A noticeable feature of equations (\ref{eq12}) and (\ref{eq13}) is that 
for  $k\ne 0$ they 
are singular at the critical values of $\beta$ 
($\beta_c$) at which $Y(\beta_c)=\pm X(\beta_c)$. In terms of the 
original  variables we have $R(\beta_c)=0$ and  
$\phi(\beta_c)=\pm\infty$\footnote{None of the values of $\beta_c$ coincide 
with the junction at which $\beta =0$.}.
Any attempt in solving these equations involves handling these ``moving 
singularities," as one encounters them when integrating the coupled equations. 
To handle these singularities, we establish jump conditions across them   as 
follows: close to $\beta=\beta_c$, we assume that the solutions have the 
following linear forms
$$ X_{\pm}=a_{\pm}+b_\pm\beta, $$
$$Y_{\pm}=c_\pm+d_\pm\beta, $$
where $\pm$ refers to the right or left hand sides of the singularity 
respectively. Substituting the above equations in (\ref{eq12}) and 
(\ref{eq13}) and integrating in the interval $\beta_c-\epsilon,\hspace{1mm}
\beta_c+\epsilon$, where $2\epsilon$ is the distance across the jump, gives 
the following equation  at $ Y_c=\pm X_c$ 
\begin{eqnarray}
b_{+}-b_{-}=-\frac{9}{4}k\frac{(2X_c\epsilon)^{1/3}}{(b_{-}\mp d_{-})^{2/3}}
\beta_c=\mp(d_{+}-d_{-}), \label{eq17}
\end{eqnarray}
where $\epsilon$ can be taken as small a value as is desired for any required 
accuracy. Moreover, the values of $a_{\pm}$ and $c_{\pm}$ are determined
by the requirement of the continuity of $X$ and $Y$ at $\beta_c$.
Equation (\ref{eq17}) establishes our jump condition for handling 
the singularities of the differential equations. For ease of comparison with 
the solutions  of the spatially flat ($k=0$) Robertson-Walker universe, we use 
the same set of parameters as in \cite{Dereli} ($b=2$, $\lambda=0$, $m^2=4.5$) 
and choose our initial conditions according to equations (\ref{eq15}) and 
(\ref{eq16}). We note that since equation (\ref{eq14}) is 
a constant of motion, if it is satisfied at $\beta=0$, it will be satisfied 
at all other values of $\beta$.  For integrating equations (\ref{eq12}) and 
(\ref{eq13}), we have used the fourth order Runge-Kutta method. The resulting 
solutions for $k=\pm1$ are shown in figure  2. As a measure of the accuracy 
of the solutions, we have  computed the ``zero energy condition,"  equation 
(\ref{eq14}), as a function of $\beta$ for $k=\pm1$. We have found that the 
values of ``total energy" stay very close to zero in the full range of 
$\beta$ shown in figure 2, thus indicating  the 
validity of our numerical solutions. As a further check, we have numerically 
recovered the analytic solutions presented in \cite{Dereli} for $k=0$. 
In figure 3 the variations of $\phi$ and $R$ and in figure 4 that of the 
curvature scalar are shown as a function of $\beta$. 

\section{Conclusions}
We have shown that in the context of classical general relativity, using a 
model proposed in \cite{Dereli}, there could exist signature transitions in 
Robertson-Walker cosmologies with arbitrary curvature. In this model, 
these transitions are embodied in the form of smooth continuous 
functions describing solutions to Einstein field equations coupled to a 
scalar field. These solutions describe a geometry in 
which the covariant metric tensor is degenerate on a hypersurface 
characterised by $\beta=0$ and undergoes a transition from a Euclidean to a 
Lorentzian signature. As in the $k=0$ case, the singular behaviour of the 
scalar 
field characterizes the beginning of the Euclidean domain. This feature  is 
also reflected in the behaviour of the curvature  scalar.  For $k=0$ the 
solutions for $X$ and $Y$  are not singular anywhere in the Lorentzian domain. 
However, for $k\ne 0$, there are  singularities at $Y=\pm X$ but they are very 
mild and show up only at the second derivative level. As $k$ increases we 
observe more rapid oscillations in $X$ and $Y$, leading to  more rapid 
variations in  all the physical quantities. However, for all values of $k$, 
they  vary extremely slowly in the vicinity of the signature changing 
hypersurface, as can be seen from figures 3 and 4. 

As far as the controversy on the junction conditions in the current 
literature is concerned, we find that in our continuous approach to 
signature change (with a continuous metric) the restrictions on the initial  
conditions are given by equations (\ref{eq15},\ref{eqdot},\ref{eqddot}). 
These conditions are slightly more restrictive than those of \cite{Hayward},
because equation (\ref{eq15}) implies that the second derivatives of $R$ and 
$\phi$ with respect to the evolution parameter $(\beta)$ are zero, whereas 
the analysis presented there does not seem to require this. 
However, the junction conditions of \cite{kk} seem to be too restrictive 
since it requires the second (cosmic) time derivatives of $R$ and $\phi$ 
to be zero, contrary to equation (\ref{eqddot}). As was mentioned before,
we only require the second derivatives with respect to $\beta$ to be zero. 

We now discuss the physical significance of the restrictions on the 
parameters of the model which is presented in equation (\ref{eq1666}). When 
this equation is written in terms of the physical parameters it becomes
\begin{eqnarray}
\frac{3\lambda}{4}-\frac{9k}{4R^2(0)}\le|b|\le \frac{m^2}{2}.
\label{eq2ineq}
\end{eqnarray}
In particular we have $\lambda_{\mbox{\footnotesize max}} = 3k/R^2(0)+2m^2/3$ 
and this implies that $\lambda$ cannot have an arbitrarily large positive 
value. Rather more importantly, 
note that as $m \rightarrow 0$, equation (\ref{eq2ineq}) implies that 
$b=0$ and subsequently equation (\ref{eq166}) shows that $\phi (0)
\rightarrow \pm \infty$. Therefore, in order to have continuous solutions 
across the junction, the scalar field has to be massive. We expect this 
result to be independent of the particular choice of the potentual $U(\phi)$.

It is  interesting to note that although  the solutions for $k=0$ and $k=\pm1$ 
are qualitatively not very different, there is a pronounced difference in the 
curvature scalar. That is ${\cal R}(0)=6k/R^2(0)$.
\vskip20pt\noindent
{\large{\bf Acknowledgements}}
\vskip8pt\noindent
We would like to thank R. W. Tucker for motivating us to do this problem and 
F. Ebrahimi for useful discussions.

\pagebreak
{\Large\bf Figure Captions}
\vspace{10mm}\\
{\bf Figure 1:} The contour plots of the allowed values of $X$ and $Y$, 
satisfying the equation of constraint (\ref{eq16}) at $\beta=0$ for 
$k=\pm1$ --- figures 1(a), 1(b).  The point (0,0) is  a solution and the 
curves approaching this point actually pass through it, although this is not 
shown on the plots due to the limitations on the numerical accuracy.
\vspace{5mm}\\
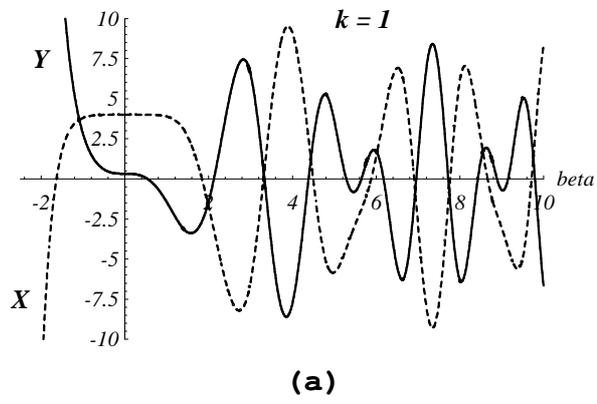
{\bf Figure 2:} Solutions for $X$ (broken curve) and $Y$ (solid curve) as a 
function of $\beta$, for $k=\pm1$ --- figures 2(a), 2(b). The values of the 
parameters are $b=2$, $\lambda=0$, $m^2=4.5$.\vspace{5mm}\\
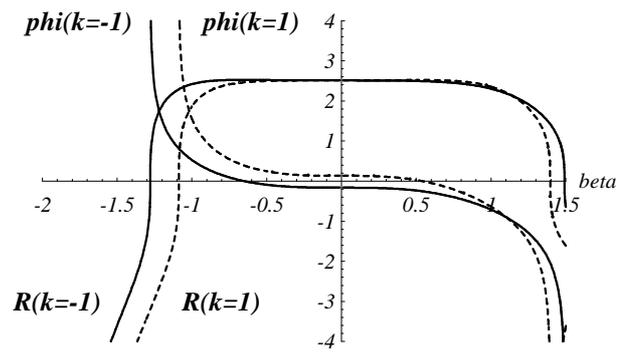
{\bf Figure 3:} Variation of $\phi$ and $R$ with $\beta$ for $b=2$, 
$\lambda=0$, $m^2=4.5$, and $k=1$ (broken curves) and $k=-1$ (solid curves). 
Note that $\phi$ and $R$ vary extremely slowly in the vicinity of 
$\beta =0.$  \vspace{5mm}\\
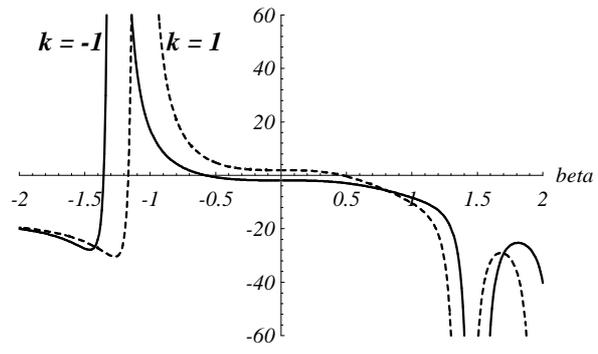
{\bf Figure 4:} Variation of the scalar curvature $\cal R$ with $\beta$ for
 $b=2$, $\lambda=0$, $m^2=4.5$, and $k=1$ (broken curve) and $k=-1$ 
(solid curve). Note that ${\cal R} (0)=6k/[R^2(0)]. $

\begin{thebibliography}{99}
\bibitem{Dewitt}B. S. DeWitt, {\it Phys. Rev.} {\bf 160}, 1113 
   (1967); C. W. Misner, {\it Phys. Rev.} {\bf 186}, 1319(1969);
W. F. Blyth and C. J. Isham, {\it Phys. Rev.} {\bf D 11}, 768 (1975).
\bibitem{Hartle} J. B. Hartle and S. W. Hawking, {\it Phys. Rev.}
{\bf D 28}, 2960 (1983);
S. W. Hawking,  {\it Nucl. Phys.}  {\bf B239}, 257 (1984);
S. W. Hawking,  300 years of gravitation ed.  S. W.  Hawking and W. Israel, 
Cambridge University Press  631 (1987);
J. J. Halliwell and J. B. Hartle,  {\it Phys. Rev.} {\bf D 41}, 1815 (1990);
G. W. Gibbons and J. B. Hartle, {\it Phys. Rev.} {\bf D 42}, 2458 (1990);
R. Kerner and J. Martin, {\it Class. Quantum Grav.} {\bf 10}, 2111 (1993);
J. Martin, {\it Phys. Rev.} {\bf D 49}, 5086 (1994).
\bibitem{kundt} W. Kundt,  {\it Commun. Math. Phys.} {\bf 4}, 143 (1966);
R. P. Geroch, {\it J. Math. Phys.} {\bf 8}, 782 (1967);
D. Brill, Magic without magic, ed. J. K. Louder (San Fransisco: Freeman)  309 
(1972).
\bibitem{HM} S. W. Hawking and I. G. Moss, {\it Phys. Lett.} {\bf 110B}, 
35 (1982); A. Vilenkin,  {\it Phys. lett.} {\bf 117B}, 25 (1982);
A. Vilenkin,  {\it Phys. Rev.} {\bf D 27}, 2848 (1983);
A. D. Linde,  {\it Phys. Lett.} {\bf 162B},  281 (1985);
A. Vilenkin,  {\it Phys. Rev.} {\bf D 33}, 3560 (1986);
I. Moss and S. Poletti,  {\it Nucl. Phys.}  {\bf B341}, 155 (1990).
\bibitem{Sakharov} A. D. Sakharov, {\it Sov. Phys.-JETP} {\bf 60}, 214 (1984);
G. F. R. Ellis, A. Sumruk, D. Coule and C. Hellaby, {\it Class. Quantum Grav.} 
{\bf 9}, 1535 (1992).
\bibitem{kkd} M. Kossowski and M. Kriele, {\it Class. Quantum Grav.} {\bf 10},
2363 (1993); T. Dray, {\it J. Math. Phys.} {\bf 37}, 5267 (1996).
\bibitem{Dereli} T. Dereli and R. W. Tucker, {\it Class. Quantum Grav.}
 {\bf 10}, 365 (1993).
\bibitem{Hayward} S. A. Hayward, {\it Class. Quantum Grav.} {\bf 9}, 
1851 (1992); {\it ibid} 2453(E);
S. A. Hayward, {\it Class. Quantum Grav.} {\bf 10}, L7 (1993).
\bibitem{sepangi} F. Darabi and H. R. Sepangi, {\it Class. Quantum Grav.} 
{\bf 16}, 1565 (1999).
\bibitem{Heldray} C. Hellaby and T. Dray, {\it Phys. Rev.} {\bf D 49}, 
5096 (1994); T. Dray and C. Hellaby, {\it J. Math. Phys.} {\bf 35}, 
5922 (1994). 
\bibitem{heyhel} S. A. Hayward, {\it phys. Rev.} {\bf D 52}, 7331 (1995);
C. Hellaby and T. Dray, {\it phys. Rev.} {\bf D 52}, 7333 (1995).
\bibitem{Ellis} 
G. F. R. Ellis,  {\it Gen. Rel. Grav.} {\bf 24}, 1047 (1992). 
\bibitem{NC}T. Dray, C. A. Manogue, and R. W. Tucker,  {\it Gen. Rel. Grav.} 
{\bf 23}, 967 (1991);
T. Dray, C. A. Manogue and R. W. Tucker, {\it Phys. Rev.} {\bf D 48}, 
2587 (1993); C. Hellaby and T. Dray, {\it Phys. Rev.} {\bf D 49}, 5096 (1994).
\bibitem{kk} M. Kossowski and M. Kriele {\it Proc. R. Soc. Lond.} {\bf A 446}, 
115 (1995).
\end{thebibliography}
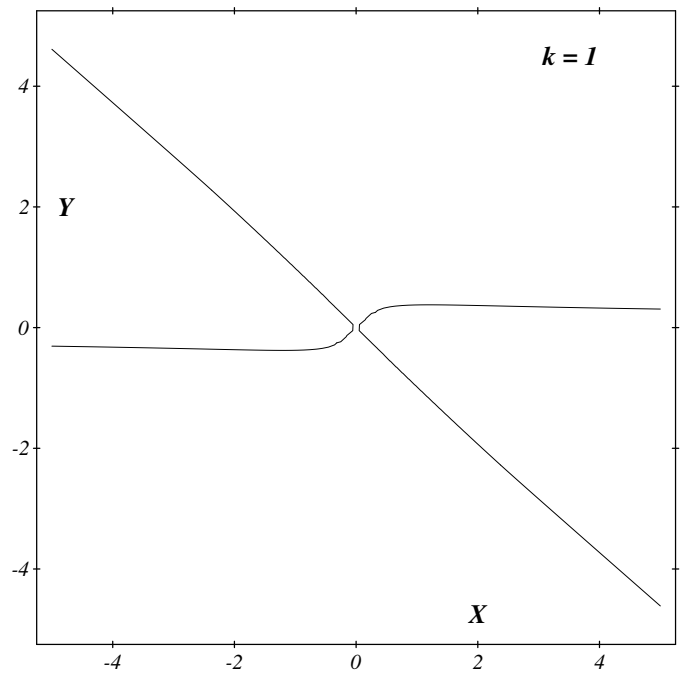
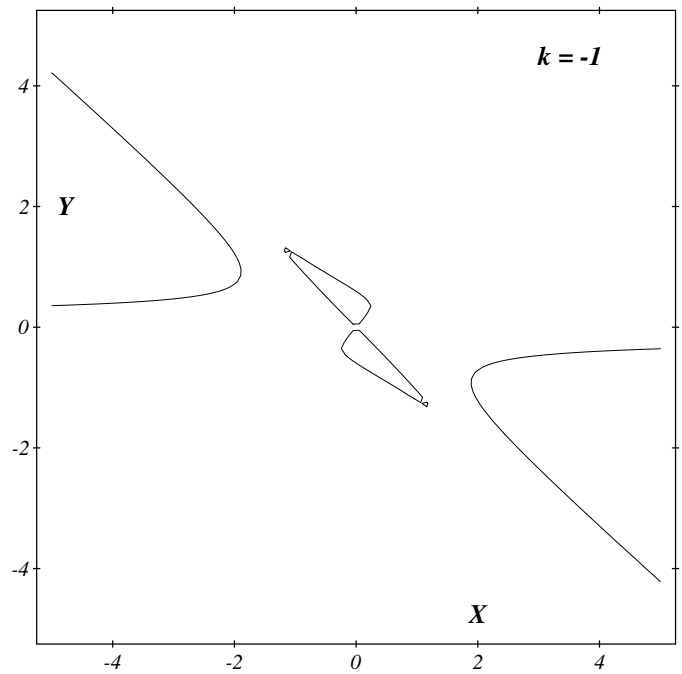
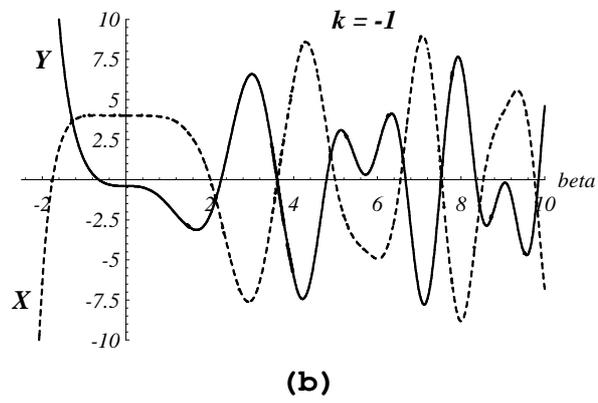
\end{document}